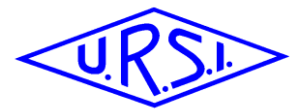

# Spectral Efficiency for mmWave Downlink with Beam Misalignment in Urban Macro Scenario

Jarosław Wojtuń[(1)], Cezary Ziółkowski[(1)], Jan M. Kelner[(1)], Aniruddha Chandra[(2)], Rajeev Shukla[(2)], Anirban Ghosh[(3)], Aleš Prokeš[(4)], Tomas Mikulasek[(4)], Radek Zavorka[(4)], and Petr Horký[(4)]

(1) Institute of Communications Systems, Military University of Technology, Warsaw, Poland
(2) Department of Electronics and Communication Engineering, National Institute of Technology, Durgapur, India
(3) Department of Electronics and Communication Engineering, SRM University AP, Andhra Pradesh, India
(4) Department of Radio Electronics, Brno University of Technology, Brno, Czech Republic

## Abstract

In this paper, we analyze the spectral efficiency for millimeter wave downlink with beam misalignment in urban macro scenario. For this purpose, we use a new approach based on the modified Shannon formula, which considers the propagation environment and antenna system coefficients. These factors are determined based on a multi-ellipsoidal propagation model. The obtained results show that under non-line-of-sight conditions, the appropriate selection of the antenna beam orientation may increase the spectral efficiency in relation to the direct line to a user.

## 1. Introduction

The spectral efficiency is based on Shannon–Hartley theorem for the additive white Gaussian noise (AWGN) channel. Generally, it can be referred to as free-space (FS) conditions or, to put it simply, also to line-of-sight (LOS) conditions [1]. Under non-LOS (NLOS) conditions, where more complex propagation conditions occur, this approach is insufficient. In addition, transmitting and receiving antenna systems should be considered in estimating the spectral efficiency. Therefore, we can conclude that the spectral efficiency value is primarily determined by the environment and antenna systems.

In this paper, we propose a spectral efficiency estimation for a millimeter wave (mmWave) downlink (DL) under beam mismatch conditions based on a modified Shannon formula that considers the coefficients of the antenna system and propagation environment. These coefficients are determined by simulation using the 3D multi-ellipsoidal propagation model (MPM) [2]. This approach is based on [3], where spectral efficiency is determined for a radio link with two directional horn antennas. This paper analyzes beam orientations of the 5G gNodeB beamforming antenna system and user equipment (UE) antenna operating in the mmWave band on the spectral efficiency. The studies included LOS and NLOS conditions, which are defined based on 3GPP tapped-delay line (TDL) model [4]. The results show a significant influence of the misalignment of antenna beams on the received power in NLOS conditions. Under LOS conditions, spectral efficiency maximization is achieved for gNodeB beam oriented to UE. Under NLOS conditions, this direction does not maximize spectral efficiency. This method of evaluating the impact of parameters and the optimal choice of antenna orientation on the radio spectral efficiency under beam mismatch conditions determines the originality and novelty of the developed solution. The main contributions in this paper are listed as follows.

- We use a novel spectral efficiency estimation methodology based on the standard Shannon formula that considers the coefficients of the antenna system and propagation environment.
- Through extensive simulation, we analyze spectral efficiency for a 28 GHz mmWave DL for urban macro (UMa) scenario using the recommended 3GPP antenna patterns.
- We put a recommendation to use optimal antenna orientation for different propagation conditions to maximize spectral efficiency.

The remainder of the paper is organized as follows. An approach to determining channel spectral efficiency is described in Section II. Section III presents the results of spectral efficiency estimation for a 28 GHz mmWave DL and UMa scenario using the MPM and recommended 3GPP antenna power radiation patterns. A summary is provided in Section IV.

## 2. Influence of Propagation Environment and Antenna System on Spectral Efficiency

The spectral efficiency $C_f$ of the radio channel depends on its quality, defined by signal-to-noise ratio, $SNR$. In the case of FS propagation, $C_f$ can be described by the Shannon's formula

$$C_f = \log_2(1 + SNR) \,[bit/s/Hz] \qquad (1)$$

where $SNR = P_f(D)/P_n$, $P_f(D)$ and $P_n$ are the desired signal power at a distance $D$ from the transmitter and noise (interference) power, respectively.

Differences in propagation conditions in a real multipath (MP) environment and the ability to concentrate energy radiation by the antenna system significantly affect the power level $P_s(D)$ of the received signal. This means that

This research was funded in part by the National Science Center (NCN), Poland, grant no. 2021/43/I/ST7/03294 (MubaMilWave). For this purpose of Open Access, the author has applied a CC-BY public copyright license to any Author Accepted Manuscript (AAM) version arising from this submission.

the spectral efficiency $C_0$ is determined by both the environmental conditions and the pattern and orientation of the antenna system. The radio wave propagation conditions in the real environment and the antenna system parameters are the reason for the different power values $P_f(D)$ and $P_s(D)$ i.e., $SNR$ and $SNR_0$, respectively.
Let $P_m(D)$ be the desired signal power for an omnidirectional antenna and MP environment. Therefore, we can express $SNR$ and $SNR_0$ by the following relation

$$SNR_0 = \frac{P_s(D)}{P_n} = \frac{P_s(D)}{P_m(D)} \cdot \frac{P_m(D)}{P_f(D)} \cdot \frac{P_f(D)}{P_n} = \frac{P_s(D)}{P_m(D)} \cdot \frac{P_m(D)}{P_f(D)} \cdot SNR \quad (2)$$

The expression describing the spectral efficiency $C_0$ considering the real propagation conditions and influence of the antenna system is [3]

$$C_0 = \log_2(1 + K_a(D)K_e(D) \cdot SNR)\,[\text{bit/s/Hz}] \quad (3)$$

where $K_a(D) = P_s(D)/P_m(D)$ and $K_e(D) = \frac{P_m(D)}{P_f(D)} = \frac{PL_f(D)}{PL_m(D)}$ represents antenna system and propagation environment factors, respectively, $PL_f(D)$ and $PL_m(D)$ are defined by path loss models for FS and MP conditions (e.g., 3GPP, WINNER II, COST 2100, or MiWEBA), respectively.
Equation (3) allows for a comparative assessment of the impact of both the propagation environment and antenna system on the radio spectral efficiency.
The main problem of the practical use of (3) to determine the spectral efficiency boils down to the evaluation of the coefficient $K_a(D)$, i.e., the relationship between the received powers using the beamforming and omnidirectional antenna systems. The MPM is used to determine the received signal power in MP propagation conditions. Figure 1 shows the MPM geometry [2]. A detailed description of the MPM geometrical structure has been presented in [2,3,5].

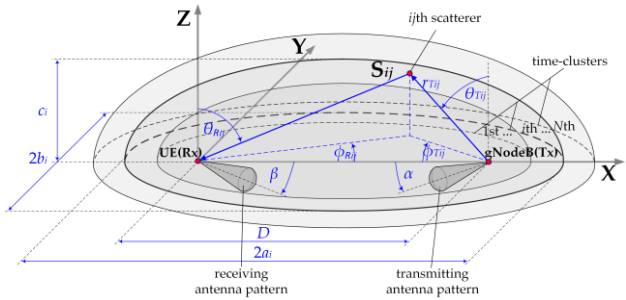

**Figure 1.** Scattering geometry of MPM.

The MPM-based methodology described in [5] modifies the path loss and power balance for different HPBWs and orientations of the antenna beams. A relative power factor, $K$, is its basis. It represents a relative power for the analyzed beam mismatch and alignment conditions, as follows

$$K(\alpha, \beta, D)[\text{dB}] = 10\log\frac{P_s(\alpha,\beta,D)}{P_s(\alpha=180°,\beta=0°,D)} = 10\log\frac{K_a(\alpha,\beta,D)}{K_a(\alpha=180°,\beta=0°,D)} \quad (4)$$

where $P_s(D) \to P_s(\alpha, \beta, D)$ is the received power for the $\alpha$ and $\beta$ directions of the transmitting and receiving antenna beams (determined with respect to the OX axe in Figure 1), respectively, and the selected distance $D$.

## 3. Spectral Efficiency Estimation for mmWave DL with Beam Misalignment in Urban Macro Scenario

The evaluation of the power losses resulting from the misalignment of the antenna beams in the directional link and the optimal selection of their orientation in LOS and NLOS conditions is based on the simulation tests. In the paper, all presented simulation studies were performed based on the MPM implementation prepared in the MATLAB environment.
In simulation studies, a spatial scenario was analyzed, as shown in Figure 1, where the 5G NR gNodeB and UE represent the transmitter (TX) and receiver (RX), respectively. Therefore, the adopted scenario may correspond to communications between the gNodeB and UE operating in the mmWave band and UMa scenario.
To model the antenna power radiation patterns, we adopted 3GPP recommendations [6]. HPBWs of the main-lobes of the antenna beams were 90° for the UE and about 12° for the gNodeB, respectively. Single antenna beam patterns of the UE and gNodeB for direction $\Phi_0 = 0°$, $\Phi_0 = 15°$ and $\Phi_0 = 30°$ are illustrated in Figure 2. In the gNodeB, we used a vertical patch as an antenna array with a size of 12×8 elements, whereas the UE antenna consists of a single element. In the rest of the paper, we will refer to these antennas as directional.

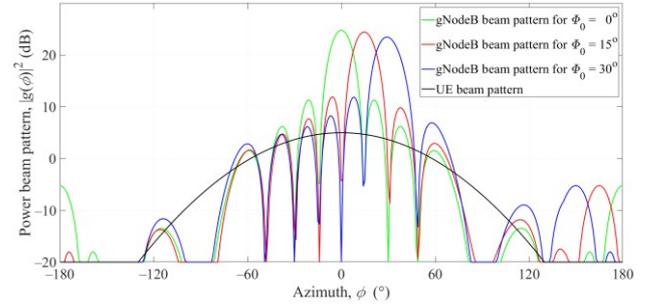

**Figure 2.** Antenna beam patterns of 1×1 UE and 12×8 gNodeB for $\Phi_0 = 0°$, $\Phi_0 = 15°$ and $\Phi_0 = 30°$.

Other assumptions are as follows:
- carrier frequency is equal to $f_c$ = 28 GHz;
- PDPs are based on TDL models from the 3GPP TR 38.901 standard [4], i.e., the TDL-B and TDL-D for NLOS and LOS conditions, respectively; these TDLs are adopted for analyzed $f_c$ and RMS delay spread, $\sigma_\tau$, for so-called the normal-delay profile and UMa scenario, i.e., $\sigma_\tau$ = 266 ns;
- Rician factor defining the direct path component in the scenario for LOS conditions is appropriate for TDL-D [4], i.e., $\kappa$ = 13.3 dB;
- distance between the TX and RX is equal to 50 m ≤ $D$ ≤ 250 m with step $\Delta D$ = 25 m;

- gains of the transmitting and receiving antennas are equal to $G_T = 25.68$ dBi and $G_R = 3.75$ dBi for the transmitting and receiving antennas, respectively;
- low heights of the transmitting (7 m) and receiving (1.5 m) antennas are based on measurement scenarios [7];
- beam alignment is defined for $\alpha = 180°$ and $\beta = 0°$ (see Figure 1);
- analyzed ranges of beam directions are as follows: $90° \leq \alpha \leq 270°$ and $-90° \leq \beta \leq 90°$ with step $\Delta\alpha = \Delta\beta = 1°$;
- to obtain average statistical results in the MPM, $L = 10$ paths are generated at the TX for each time-cluster (semi-ellipsoid); on the other hand, $M = 360$ Monte-Carlo simulations were run for each analyzed scenario; in this case, the average resolution of generating the angles of departure is about 0.1°.

In Figure 3, $K(\alpha, \beta)$ versus $\alpha$ and $\beta$ directions of the transmitting and receiving antenna beams is illustrated. This graphs clearly show that when the beams are directed at each other, a dominant received power is obtained. In the analyzed LOS and NLOS conditions. Figure 3 (a) shows that the maximum power is obtained for beam alignment, which is obvious. The direction of the receiving antenna has a decisive influence on the power level. Despite the direction changes of the TX antenna, the extremum power is ensured when the RX antenna is pointed at the TX.

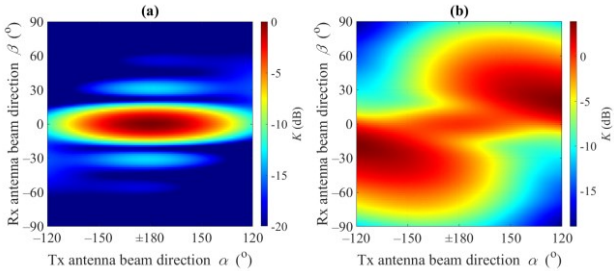

**Figure 3.** Relative power factor $K(\alpha,\beta)$ versus $\alpha$ and $\beta$ directions of transmitting and receiving beams under (a) LOS and (b) NLOS conditions for $D = 100$ m.

Under NLOS conditions, the MP propagation phenomenon makes it necessary to search for optimal $\alpha$ and $\beta$ directions of the transmitting and receiving antenna beams, which will ensure the maximization of the received signal level. Figure 3 (b) shows that under NLOS conditions, the received signal obtains the statistically highest power level for the TX beam direction equal to $\alpha = \pm 120°$. However, in this case, the beam direction of the receiving antenna to achieve this power level should be equal $\beta = \pm 28°$.

The lack of the direct path ($\kappa = 0$ dB) under NLOS conditions is the principal cause of the difference in results concerning those obtained for LOS conditions. For NLOS conditions, most of the received power comes from the delayed components scattered on the semi-ellipsoids. Therefore, the global maximum of the received power does not appear for $\alpha = \pm 180°$ and $\beta = 0°$.

The effects of using directional antenna systems are shown in Figures 4 and 5.

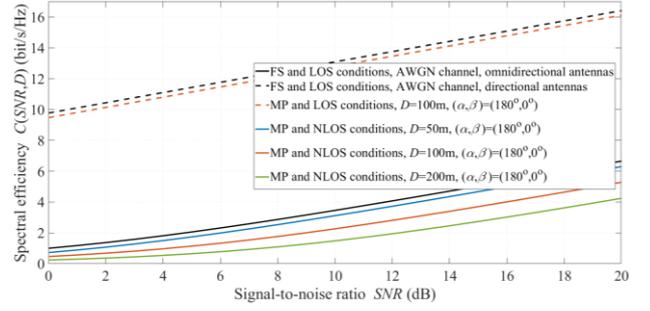

**Figure 4.** Spectral efficiency versus SNR for directional antenna pattern and under LOS and NLOS conditions.

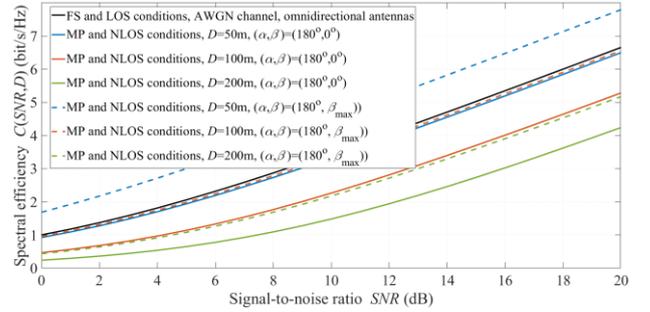

**Figure 5.** Spectral efficiency versus SNR for $\beta_{max}$ under NLOS conditions.

Under LOS conditions, the use of directional antennas and alignment of the transmitting and receiving antenna beams (i.e., beams oriented to each other) provides statistically multiple increases in the radio spectral efficiency (see Figure 4). Of course, this increase depends on the gains of the antennas. For the analyzed radio link with the directional antennas whose gains are equal to $G_T = 25.68$ dBi and $G_R = 3.75$ dBi, the spectral efficiency is higher by 9.5 bit/s/Hz concerning the radio link with the omnidirectional antenna systems and FS propagation conditions. On the other hand, for LOS conditions, MP propagation has a negligible effect on spectral efficiency compared to FS propagation.

The NLOS conditions significantly reduce the spectral efficiency even several times. The use of beamforming antenna system is one way to minimize the adverse effects of MP propagation under NLOS conditions.

The analysis results presented in Figures 4 and 5 show that the radio spectral efficiency also depends on the distance between the TX and RX. For NLOS conditions, the double distance reduction increases the radio spectral efficiency by about 0.2–1.2 bit/s/Hz in the whole analyzed range of SNR variability.

Under NLOS conditions, the alignment of the transmitting and receiving antenna beams does not provide to achieve the maximum received power. Therefore, under these propagation conditions, beamforming antenna system should supply a beam steering mechanism to the direction of the maximum level of the received signal. The justification for applying such a solution is illustrated in Figure 5.

It can be seen that using the direction of the maximum signal level ensures an additional increase in the spectral efficiency by 1 bit/s/Hz. The increment increases as the TX–RX distance is greater.

The spectral efficiency change versus the distance between the TX and RX. The comparison of the spectral efficiency change for the straight (i.e., $\alpha = \pm 180°$ and $\beta = 0°$) and optimal (i.e., $\alpha = \pm 180°$ and $\beta = \beta_{max}$) directions of the antenna beams is shown in Figure 6.

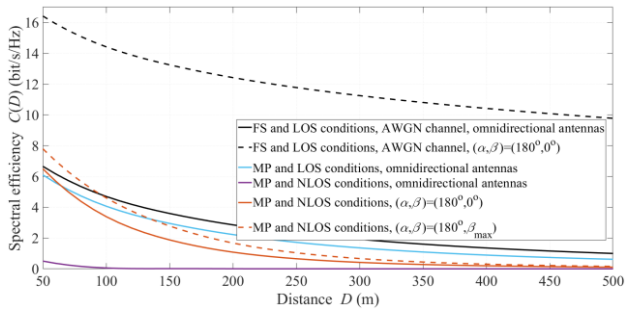

**Figure 6.** Spectral efficiency versus TX–RX distance for straight and optimal directions of antenna beams under NLOS conditions.

The exemplary graphs are obtained assuming that the received signal level provides the *SNR* = 20 dB. It is evident that as the TX–RX distance increases, the level of the desired signal decreases. Thus, the radio spectral efficiency decreases. In the case of the optimal direction of the antenna beams, a 4-fold increase in the distance (from 50 m to 200 m) causes only about a 4.5-fold reduction in the spectral efficiency. On the other hand, with the increase in the distance, maintaining the straight direction results in a 6-fold decrease in the spectral efficiency. It shows that using steering and selecting the optimal beam direction in the antenna system mitigates the degrading effect of distance on the radio spectral efficiency.

## 4. Conclusions

This paper focuses on spectral efficiency analysis for mmWave band and UMa scenario. The spectral efficiency evaluation is based on a modified Shannon formula that considers the coefficients of the antenna systems and the propagation environment. The MPM was used to determine these coefficients. Pointing the antennas at each other under LOS conditions maximizes spectral efficiency. However, the results obtained for NLOS conditions show that the optimal selection of the gNodeB beam direction can increase the spectral efficiency in relation to the direction toward the user. These optimal directions are determined based on the MPM. The conducted analysis shows that beam misalignment can positively affect the parameters of the radio link under NLOS conditions, which is in accordance with the measurement results presented in the literature.

## 6. Acknowledgements

This work was co-funded by the Czech Science Foundation under grant no. 23-04304L, the National Science Centre, Poland, under the OPUS call in the Weave program, under research project no. 2021/43/I/ST7/03294 acronym 'MubaMilWave' and by the Military University of Technology under grant no. UGB/22-863/2023/WAT, and chip-to-startup (C2S) program no. EE-9/2/2021-R&D-E sponsored by MeitY, Government of India.